\documentclass[12pt]{article}

\usepackage{a4wide}
\usepackage{amsmath,amstext}
\usepackage{amsfonts}
\usepackage{amsthm}

\newtheorem{theorem}{Theorem}
\newtheorem{proposition}[theorem]{Proposition}
\newtheorem{lemma}[theorem]{Lemma}


\begin{document}

\baselineskip 21pt

\parskip 7pt

\hfill May 1995

\vspace{24pt}

\begin{center}


  \textbf{\large
    Simple Construction  of
    Elliptic Boundary $ K $-matrix
    \footnote[5]{~arch-ive/9507123}
    }

  \vspace{24pt}
  Kazuhiro \textsc{Hikami}
  \footnote[2]{\texttt{
      hikami@monet.phys.s.u-tokyo.ac.jp}}

  \vspace{8pt}

  \textsl{Department of Physics, Graduate School of Science,} \\
  \textsl{University of Tokyo,} \\
  \textsl{Hongo 7-3-1, Bunkyo, Tokyo 113, Japan.}

  \vspace{8pt}


(Received: June 6, 1995)

\end{center}

\vspace{18pt}

\begin{center}
  \underline{ABSTRACT}
\end{center}


We give the infinite-dimensional representation for
the elliptic $ K $-operator satisfying the boundary
Yang-Baxter equation.
By restricting the functional space to  finite-dimensional space,
we construct  the elliptic   $ K $-matrix
associated to  Belavin's
completely  $ \mathbb{Z} $-symmetric
$ R $-matrix.


\vfill

\noindent
PACS:


\newpage

The quantum $ R $-matrix as  solutions  of the Yang-Baxter equation
(YBE)  has
received much attention  in  mathematical physics.
The algebraic
structure  reveals as quantum group.
Recently
the  $ R $-matrix has
been treated as operator acting on functional space.
In one sense this gives the infinite-dimensional representation for
solutions of YBE.
By use of operator description  for $ R $-matrix,
the construction of $ R $-matrix, especially for elliptic case,
becomes much simpler.
Based on the elliptic $ R $-operator
defined by Shibukawa and Ueno~\cite{ShiUen92},
Felder and Pasquier
constructed   Belavin's elliptic $ R $-matrix~\cite{FelPas94}.

The $ R $-matrix has been  used to study spin chains with
periodic boundary condition in terms of the quantum inverse scattering
method.
Besides the $ R $-matrix, the other matrix called $ K $-matrix is
used to solve the spin chain with open boundary~\cite{Chere84,Skl88}.
In this letter
we propose a method to
construct the boundary $ K $-matrix associated with
Belavin's $ R $-matrix.

Throughout this paper  we use the doubly periodic   function
$ \sigma_\mu (z) \equiv  \sigma_\mu (z,\tau) $,
\begin{equation*}
  \begin{aligned}
    \sigma_\mu (z+1) & =  \sigma_\mu (z) , \\
    \sigma_\mu (z+\tau) & = e^{2 \pi \mathrm{i} \mu}
    \, \sigma_\mu (z) ,
  \end{aligned}
\end{equation*}
where $ \tau $ is an arbitrary complex  number, satisfying
$ \mathrm{Im} \, \tau > 0 $.
The function $ \sigma_\mu(z) $ only   has  simple poles  on the
lattice
$ \mathbb{Z} + \tau \mathbb{Z} $, and the residue at origin
equals to one.
Note that
the function
$ \sigma_\mu (z) $
can be explicitly written as
\begin{equation}
  \sigma_\mu (z) = \frac{\vartheta_1(z-\mu,\tau) \,
    \vartheta_1^{\prime}(0,\tau)}
  {\vartheta_1(z,\tau) \, \vartheta_1(-\mu,\tau)} ,
  \label{def_sigma}
\end{equation}
where $ \vartheta_1(z,\tau) $ is the Jacobi's theta function,
\begin{equation}
  \vartheta_1(z,\tau)
  =
  \sum_{n \in \mathbb{Z}+\frac{1}{2}}
  \exp
  \biggl(
    \mathrm{i} \pi n^2 \tau + 2 \pi \mathrm{i} n
    \Bigl( z + \frac{1}{2} \Bigr)
  \biggr) .
\end{equation}
For the elliptic function
$ \sigma_\mu (z) $,
we have  following lemma;
\begin{lemma}
  \label{lem:prop_sigma}
  The elliptic function $ \sigma_\mu(z) $
  defined in~(\ref{def_sigma})
  satisfies the following
  identities;
  \begin{enumerate}
    \renewcommand{\labelenumi}{(\alph{enumi})}
    \item
      $ \sigma_\mu(z) = - \sigma_z(\mu) $ ,
    \item
      $ \sigma_\mu(z) = - \sigma_{-\mu}(-z)  $,
    \item
      $ \sigma_\mu(z) \, \sigma_{-\mu}(z) = \wp(z) - \wp(\mu) $ ,
    \item
      $ \sigma_\lambda(z) \,  \sigma_\mu(w)
      - \sigma_{\lambda+\mu}(w)  \, \sigma_\lambda(z-w)
      - \sigma_\mu(w-z) \, \sigma_{\lambda+\mu}(z) = 0 $ ,
    \item
      $  \sigma_\lambda(z)  \, \sigma_\mu(z)
    = \sigma_{\lambda+\mu}(z)
    \cdot
    \Bigl(
      \zeta(z) - \zeta(\lambda) - \zeta(\mu) - \zeta(z-\lambda-\mu)
    \Bigr) $ .
  \end{enumerate}
\end{lemma}
\noindent
The proof can be seen in, for example, Ref.~\cite{WhittWatso27}.
Another property for the function
$ \sigma_\mu (z) $ is as follows;
\begin{lemma}[\cite{FelPas94}]
  \label{lem:sig_sig}
  \begin{equation}
    \sigma_\mu (z,\tau)
    = \frac{1}{k}
    \sum_{a=0}^{k-1} \sigma_{\frac{\mu+a}{k}}
    (z,\frac{\tau}{k}) .
  \end{equation}
\end{lemma}
\noindent
This identity is easy to be proved
when one compares periodicity and residues of both hand sides;
one can check
that the both hand sides have simple poles on
$ \mathbb{Z} + \tau \mathbb{Z} $.
This lemma becomes useful when we construct Belavin's completely
$ \mathbb{Z} $-symmetric $ R $-matrix~\cite{Bela81}.

In terms of the elliptic function
$ \sigma_\mu(z) $,
Shibukawa and Ueno introduced  the  ``infinite-dimensional''
representation for
$ R $-operator
as  a solution of YBE.
\begin{theorem}[\cite{ShiUen92}]
  \label{th:shib_ueno}
  Let $ R $-operator acts on the space of functions of two variables,
  \begin{equation}
    R(\xi) \, f(z_1,z_2)
    = \sigma_\mu(z_{12}) \, f(z_1,z_2)
    - \sigma_\xi(z_{12} ) \, f(z_2, z_1) .
    \label{ell_r_op}
  \end{equation}
  This $ R $-operator satisfies YBE,
  \begin{equation}
    R_{12}(\xi_{12}) R_{13}(\xi_{13}) R_{23}(\xi_{23})
    = R_{23}(\xi_{23}) R_{13}(\xi_{13}) R_{12}(\xi_{12}) .
  \end{equation}
\end{theorem}
\noindent
Here and hereafter we use notations,
$ z_{12} \equiv z_1 - z_2 $,
etc.
The proof follows by using Lemma~\ref{lem:prop_sigma}.
Remark that we can define rational and trigonometric
$ R $-operators
as degenerate cases of elliptic  operator~(\ref{ell_r_op});
\begin{equation}
  \sigma_\mu (z) \to
  \begin{cases}
    \displaystyle{
      \cot z - \cot  \mu ,
      }
    & \qquad \text{trigonometric}, \\
    \noalign{\vskip 3mm}
    \displaystyle{
      z^{-1} - \mu^{-1} ,
      }
    & \qquad \text{rational}.
  \end{cases}
\end{equation}
These degenerate  types of infinite-dimensional
representations for $ R $-operator were also
studied in
Ref.~\cite{Gaud88}.

To introduce the generalized   Shibukawa-Ueno's
$ R $-operator, Felder and Pasquier introduced the
gauge-transformation.
They defined  the translation operator in functional space as
\begin{equation}
  T_k(\xi) \, f(z) = f(z - \frac{\xi}{k}) .
\end{equation}
For the translation operator $ T_k(\xi) $, we have following
identities;
\begin{lemma}
  \label{lem:r_com_tt}
  \begin{enumerate}
    \renewcommand{\labelenumi}{(\alph{enumi})}
    \item
      $ T_k(\xi+\eta) = T_k(\xi) \, T_k(\eta) $,
    \item
      $ [ R(\xi) , T_k(\eta) \otimes T_k(\eta) ] = 0 $.
  \end{enumerate}
\end{lemma}
\noindent
The first identity is trivial.
The second one  is due to the fact that $ R $-operator
depends only on the difference of two  spectral parameters.

By use of the translation operator $ T_k $, we can introduce the
`modified' $ R $-operator as a solution of  YBE.
\begin{theorem}[\cite{FelPas94}]
  \label{th:mod_r}
  Let the modified $ R $-operator be
  \begin{equation}
    R_k(\xi_{12})
    =
    \Bigl( T_k(\xi_1-\mu)^{-1} \otimes T_k(\xi_2)^{-1} \Bigr) \cdot
    R(\xi_{12})
    \cdot
    \Bigl( T_k(\xi_1) \otimes T_k(\xi_2-\mu) \Bigr) .
    \label{modified_r}
  \end{equation}
The operator
$ R_k(\xi) $
also satisfies YBE.
\end{theorem}
\noindent
Remark that the action of the modified $ R $-operator
$ R_k(\xi) $
on functional space is
explicitly written as
\begin{equation}
  \label{explicit_mod_r}
  \begin{split}
    R_k(\xi) \, f(z_1,z_2)
    & =  \sigma_\mu ( z_{12} + \frac{\mu + \xi}{k} ) \,
    f(z_1+\frac{\mu}{k},z_2-\frac{\mu}{k})  \\
    & \qquad
    - \sigma_\xi (z_{12} + \frac{\mu+\xi}{k} ) \,
    f(z_2 - \frac{\xi}{k},z_1+\frac{\xi}{k}) .
  \end{split}
\end{equation}
This modified $ R $-operator $ R_k(\xi) $
plays a crucial role in
defining Belavin's
completely $ \mathbb{Z} $-symmetric $ R $-matrix.

On the other hand,
we shall introduce the elliptic boundary $ K $-operator.
The  $ K $-matrix
was originally
introduced to solve the spin system with open
boundary based on the quantum inverse scattering
method;
algebraically $ K $-matrix satisfies the so-called
boundary YBE
(reflection equation)~\cite{Chere84,Skl88}.
Here
we  regard the $ K $-matrix as an operator acting on the
functional space.
\begin{theorem}
  Let  boundary $ K $-operator act on the space of functions of single
  variable,
  \begin{subequations}
    \begin{align}
      K^{\mathrm{I}} (\xi) \, f(z)
      & =  \sigma_{2 \xi} (z) \, f(z)
      -  \sigma_{2 \nu} (z) \, f(-z)  , \\
      K^{\mathrm{II}} (\xi) \, f(z)
      & =  \sigma_{\xi} (2z) \, f(z)
      - \sigma_{\nu} (2 z) \, f(-z) .
    \end{align}
  \end{subequations}
  These $ K $-operators satisfy  the boundary YBE,
  \begin{equation}
    \begin{split}
      &  R_{21}(\xi_1-\xi_2) \,
      \Bigl( K(\xi_1) \otimes 1 \Bigr) \,
      R_{12}(\xi_1+\xi_2) \,
      \Bigl( 1 \otimes K(\xi_2) \Bigr) \\
      & \quad
      = \Bigl( 1 \otimes K(\xi_2) \Bigr) \,
      R_{21}(\xi_1+\xi_2) \,
      \Bigl( K(\xi_1) \otimes 1 \Bigr) \,
      R_{12}(\xi_1-\xi_2) ,
    \end{split}
    \label{bybe}
  \end{equation}
  where $ R $-operator is defined in Theorem~\ref{th:shib_ueno}.
\end{theorem}

This theorem can be proved as follows.
We  suppose the action of  the  $ K $-operator as
\[
  K(\xi) \, f(z) = G(\xi,z) \, f(z) - H(z) \, f(-z) .
\]
Substituting the definition of $ R $- and $ K $-operators
into the boundary YBE~(\ref{bybe}),
we obtain   three functional equations~\cite{Hikam95d};
\begin{subequations}
  \begin{equation}
    \begin{split}
      & G(\xi_2,z_2) \, \sigma_{\xi_1+\xi_2}(z_1-z_2) \,
      \sigma_{\xi_1-\xi_2}(z_1+z_2)
      +
      G(\xi_1,z_1) \, \sigma_{\mu}(z_1+z_2) \, \sigma_{\mu}(-z_1-z_2)
      \\
      & \qquad +
      G(\xi_1,-z_2) \, \sigma_{\xi_1+\xi_2}(z_1+z_2) \,
      \sigma_{\xi_1-\xi_2}(z_1+z_2) \\
      & \quad =  G(\xi_1,z_1) \,
      \sigma_\mu(z_1-z_2) \, \sigma_\mu(z_2-z_1)
      + G(\xi_1,z_2) \,  \sigma_{\xi_1-\xi_2}(z_1-z_2)  \,
      \sigma_{\xi_1+\xi_2}(z_1-z_2)  \\
      & \qquad +
      G(\xi_2,-z_2)  \,  \sigma_{\xi_1-\xi_2}(z_1-z_2) \,
      \sigma_{\xi_1+\xi_2}(z_1+z_2) ,
    \end{split}
  \end{equation}
%
%
  \begin{equation}
    \begin{split}
      & G(\xi_1,z_1) \, G(\xi_2,z_2) \,
      \sigma_{\xi_1-\xi_2}(z_2-z_1)
      + G(\xi_1,z_2) \, G(\xi_2,z_2) \,
      \sigma_{\xi_1+\xi_2}(z_1-z_2)
      \\
      & \qquad +
      H(-z_2) \, H(z_2) \,
      \sigma_{\xi_1+\xi_2}(z_1+z_2) \\
      & \quad =  G(\xi_1,z_1) \, G(\xi_2,z_1) \,
      \sigma_{\xi_1+\xi_2}(z_2-z_1)
      + G(\xi_1,z_2) \, G(\xi_2,z_1) \,
      \sigma_{\xi_1-\xi_2}(z_1-z_2)
      \\
      & \qquad +
      H(z_1) \, H(-z_1) \, \sigma_{\xi_1+\xi_2}(z_1+z_2) ,
    \end{split}
  \end{equation}
%
%
%
  \begin{equation}
    \begin{split}
      & G(\xi_1,z_1) \, \sigma_{\xi_1+\xi_2}(z_2-z_1)
      + G(\xi_1,z_2) \, \sigma_{\xi_1-\xi_2}(z_1-z_2)
      + G(\xi_2,-z_1) \, \sigma_{\xi_1+\xi_2}(z_1+z_2)
      \\
      & \quad =  G(\xi_2,z_2) \, \sigma_{\xi_1-\xi_2}(z_1+z_2) .
    \end{split}
  \end{equation}
\end{subequations}
One can conclude by comparing the periodicity and
residues of the both hand sides that functions
\[
  \begin{array}{rl}
    \mathrm{I}. & G(\xi,z) = \sigma_{2 \xi}(z) ,
    \qquad
    H(z) = \sigma_{2 \nu}(z) , \\
    \noalign{\vskip 3mm}
    \mathrm{II}. & G(\xi,z) = \sigma_{\xi}(2 z) ,
    \qquad
    H(z) = \sigma_{\nu}(2 z) ,
  \end{array}
\]
solve  these  functional equations.
In this calculation, we have
used Lemma~\ref{lem:prop_sigma}.

We  also introduce the modified $ K $-operator associated with
modified $ R $-operator as a solution of the boundary
YBE~(\ref{bybe}).
\begin{theorem}
  Let the modified $ K $-operators be
  \begin{equation}
    K_k^{\mathrm{I, II}} (\xi)
    = T_k(\xi + \nu) \, K^{\mathrm{I,II}} (\xi) \, T_k(\xi - \nu) .
  \end{equation}
  The modified operators
  $ K_k^{\mathrm{I, II}} (\xi) $ and $ R_k(\xi) $ also satisfy
  the boundary YBE~(\ref{bybe}).
\end{theorem}
\noindent
This Theorem follows with help of  Lemma~\ref{lem:r_com_tt}.
Note that explicit forms of action of the modified $ K $-operator
can be  written as
\begin{subequations}
  \label{explicit_mod_k}
  \begin{align}
    K_k^{\mathrm{I}} (\xi) \, f(z)
    & =
    \sigma_{2 \xi}(z + \frac{\xi + \nu}{k} ) \,
    f(z + \frac{2\xi}{k})
    - \sigma_{2 \nu} (z + \frac{\xi + \nu}{k} ) \,
    f(-z-\frac{2 \nu}{k}) , \\
    K_k^{\mathrm{II}} (\xi) \, f(z)
    & =
    \sigma_\xi (2z+\frac{2\xi + 2\nu}{k}) \,
    f(z+\frac{2\xi}{k})
    - \sigma_\nu ( 2z + \frac{2\xi + 2\nu}{k}) \,
    f(-z-\frac{2\nu}{k}) .
  \end{align}
\end{subequations}

Now by use of the modified operators, $ R_k(\xi) $ and
$ K_k^{\mathrm{I, II}} (\xi) $, we shall construct
Belavin's completely
$ \mathbb{Z} $-symmetric
$ R $-matrix~\cite{Bela81,Bax82,Chere82,Bovie83,Trac85}
and its
$ K $-matrix~\cite{InamKonn94,VegaGonz94b,FanHouShiYan95}.
This can be done by restricting functional space
into finite-dimensional space~\cite{FelPas94};
define $ V_k $ as the space of entire functions such that
\begin{equation*}
  \begin{aligned}
    f(z+1) & = f(z) , \\
    f(z+\tau) & =
    e^{-2 \pi \mathrm{i} k z - \pi \mathrm{i} k \tau }
    \,
    f(z) .
  \end{aligned}
\end{equation*}
The space $ V_k $ has dimension $ k $.
As bases of $ k $-dimensional functional space $ V_k $,
we use the
$ \theta $-function for
$ a \in \mathbb{Z}_k \equiv \mathbb{Z} / k \mathbb{Z} $
defined by
\begin{equation}
  \theta_a (z) = \sum_{n \in \mathbb{Z}}
  \exp
  \biggl(
    \pi \mathrm{i} n^2  \frac{\tau}{k}
    + 2 \pi \mathrm{i} n  \Bigl( z- \frac{a}{k} \Bigr)
  \biggr).
\end{equation}
We note that the functions $ \theta_a (z) $ have properties,
\[
  \widehat{S} \, \theta_a (z) = \theta_{a-1}(z) ,
  \qquad
  \widehat{T} \, \theta_a (z)
  = e^{2 \pi \mathrm{i} a / k} \theta_a (z) ,
\]
where operators $ \widehat{S} $ and $ \widehat{T} $ act on the
functional space as follows;
\[
  \widehat{S} \, f(z) = f(z+\frac{1}{k}) ,
  \qquad
  \widehat{T} \, f(z)
  = e^{2 \pi \mathrm{i} z + \pi \mathrm{i} \tau /k} f(z+\frac{\tau}{k}) .
\]

When one checks periodicity
of the modified $ R_k(\xi) $ and
$ K_k(\xi) $ operators in~(\ref{explicit_mod_r})
and~(\ref{explicit_mod_k})
respectively, we shall have the following proposition;
\begin{proposition}
  \begin{enumerate}
    \renewcommand{\labelenumi}{(\alph{enumi})}
  \item
    Modified operator
    $ R_k(\xi) $   preserves
    $ V_k \otimes V_k $.

  \item
    Modified operator
    $ K_k^{\mathrm{I, II}} (\xi) $     preserves
    $ V_k $.

  \end{enumerate}
\end{proposition}
\noindent
This proposition is easy to be proved by direct calculations.

Based on above proposition,
we can restrict the functional space, on which modified
$R$- and $K$-operators act,
to $ V_k $.
In this finite-dimensional functional space,
we can recover
Belavin's $ \mathbb{Z} $-symmetric solution from the modified
$ R_k $-operator.
\begin{theorem}[\cite{FelPas94}]
  Define matrix elements of modified $ R $-operator
  $ R_k(\xi) $ by
  \begin{equation}
    R_k(\xi) \, \theta_a \otimes \theta_b
    = \sum_{c, d \in \mathbb{Z}_k}
    R_k(\xi)_{a c, b d}
    \, \theta_c \otimes \theta_d .
  \end{equation}
  Then we get,
  \begin{equation}
    \label{belavin_element}
    R_k(\xi)_{a c, b d}
    = \delta_{a+b,c+d}
    \cdot
    \frac{\vartheta_1(\frac{\mu-\xi-a+b}{k},\frac{\tau}{k})
      \, \vartheta_1^{\prime}(0,\frac{\tau}{k})}
    {k \, \vartheta_1(\frac{\mu-a+c}{k},\frac{\tau}{k})
      \, \vartheta_1(\frac{\xi-b+c}{k},\frac{\tau}{k})}
  \end{equation}
\end{theorem}
\noindent
The proof should  be done by use of Lemma~\ref{lem:sig_sig}.
The key is to rewrite
$ R_k(\xi) \, \theta_a (z_1) \otimes \theta_b (z_2) $ as
\begin{equation*}
  \begin{aligned}
    \frac{1}{k} &
    \sum_{c \in \mathbb{Z}_k}
    \Bigl\{
      \sigma_{\frac{\mu+c-a}{k}}
      (z_{12} +\frac{\mu+\xi}{k},\frac{\tau}{k})
      \,
      \theta_a (z_1+\frac{\mu}{k})
      \,
      \theta_b (z_2 - \frac{\mu}{k})
      \\
    & \qquad
      - \sigma_{\frac{\xi+c-b}{k}}
      (z_{12}+\frac{\mu+\xi}{k})
      \,
      \theta_a (z_2-\frac{\xi}{k})
      \,
      \theta_b (z_1+\frac{\xi}{k})
    \Bigr\} .
  \end{aligned}
\end{equation*}
Each summand is an eigenstate of
$ \widehat{T} \otimes 1 $ with eigenvalue
$ e^{2 \pi \mathrm{i} c/k} $,
which proves that the summand is proportional to
$ \theta_c \otimes \theta_d $ with
$ d=a+b-c $.
By setting
$ z_{12} = (-\xi+c-a)/k $ and
using the fact,
$ \theta_a (z_2 - \frac{\xi}{k}) \,
\theta_b (z_1 + \frac{\xi}{k} ) =
\theta_c(z_1) \, \theta_d (z_2) $,
we can obtain above expression.
For $ k=2 $ it reduces to $R$-matrix of the Baxter's eight vertex
model (XYZ spin chain).

In the same manner, the restriction of
modified operator
$ K_k(\xi) $
into $ V_k $
gives the boundary $ K $-matrix associated with Belavin's
$R$-matrix~(\ref{belavin_element}).
\begin{theorem}
  Define matrix  elements of the modified
  $ K^{\mathrm{I}} $-operator by
  \begin{equation}
    K_k^{\mathrm{I}} (\xi) \, \theta_a (z)
    =
    \sum_{c \in \mathbb{Z}_k}
    K_k^{\mathrm{I}} (\xi)_{a, c} \, \theta_c (z)  .
  \end{equation}
  Then we get,
  \begin{equation}
    K_k^{\mathrm{I}} (\xi)_{a, c}
    =
    \frac{
      \vartheta_1(\frac{2\nu+2a-2\xi}{k},\frac{\tau}{k}) \,
      \vartheta_1^{\prime}(0,\frac{\tau}{k})
      }
    {k \, \vartheta_1(\frac{2\nu+c+a}{k},\frac{\tau}{k}) \,
      \vartheta_1(\frac{-2\xi-c+a}{k},\frac{\tau}{k})}
    \cdot
    \frac{\theta_{c} (\frac{- \nu - \xi}{k})}
    {\theta_{a} (\frac{-\nu + \xi}{k})} .
  \end{equation}
\end{theorem}
\noindent
Outline of proof is essentially same with the case of
$ R $-matrix.
Based on Lemma~\ref{lem:sig_sig},
we can rewrite $ K_k^{\mathrm{I}} (\xi) \, \theta_a (z) $
as
\[
  \frac{1}{k} \sum_{c \in \mathbb{Z}_k}
  \Bigl\{
    \sigma_{\frac{2 \xi+c-a}{k}}
    (z+\frac{\xi + \nu}{k},\frac{\tau}{k})
    \,
    \theta_a (z+\frac{2\xi}{k})
    - \sigma_{\frac{2\nu+c+a}{k}}
    (z+\frac{\xi + \nu}{k},\frac{\tau}{k})
    \,
    \theta_a (-z-\frac{2\nu}{k})
  \Bigr\} .
\]
Each summand  is an eigenstate of
$\widehat{T} $ with eigenvalue $ e^{2 \pi \mathrm{i} c /k} $,
which shows that summand is proportional to $ \theta_c (z) $
for arbitrary $ z $.
The prefactor can be calculated by setting
$ z = (- \xi + \nu + c +a)/k $.

The elliptic matrix $K_k^{\mathrm{I}} (\xi) $ coincides with result of
Ref.~\cite{FanHouShiYan95}
(for trigonometric case, Ref.~\cite{KuliSkly91}).
For $ k=2 $ case, it
gives one of the boundary $K$-matrix for the Baxter's eight vertex
model studied in Ref.~\cite{InamKonn94}.

For the second-type $ K_k^{\mathrm{II}}(\xi) $ operator,
the matrix elements can be calculated   for odd-$k$ case.
\begin{theorem}
  Define matrix elements of the modified
  $ K^{\mathrm{II}} $-operator
  $ K_k^{\mathrm{II}}(\xi) $ by
  \begin{equation}
    K_k^{\mathrm{II}} (\xi) \, \theta_{2a} (z)
    =
    \sum_{c \in \mathbb{Z}_k}
    K_k^{\mathrm{II}} (\xi)_{2a,2c}  \, \theta_{2c} (z) .
  \end{equation}
  To choose $ \theta_{2a} $ as bases of $ V_k $,
  we must assume that dimension $ k $ is odd.
  In this case, we obtain that the matrix elements have the form,
  \begin{equation}
    K_k^{\mathrm{II}} (\xi)_{2a,2c}
    =
    \frac{\vartheta_1(\frac{\nu-\xi+2a}{k},\frac{\tau}{k}) \,
      \vartheta_1^{\prime}(0,\frac{\tau}{k})}
    {k \, \vartheta_1(\frac{\nu+c+a}{k},\frac{\tau}{k}) \,
      \vartheta_1(\frac{-\xi-c+a}{k},\frac{\tau}{k})}
    \cdot
    \frac{\theta_{2a} (\frac{2\xi - \nu+c+a}{2k})}
    {\theta_{2c} (\frac{-2\xi - \nu+c+a}{2k})} .
  \end{equation}
\end{theorem}
\noindent
The key of  proof is to rewrite
$ K_k^{\mathrm{II}}(\xi) \, \theta_{2a}(z) $ as
\[
  \frac{1}{k} \sum_{c \in \mathbb{Z}_k}
  \Bigl\{
    \sigma_{\frac{\xi+c-a}{k}} (2z+\frac{2\xi+2\nu}{k}) \,
    \theta_{2a}(z+\frac{2\xi}{k})
    -
    \sigma_{\frac{\nu+c+a}{k}} (2z+\frac{2\xi+2\nu}{k}) \,
    \theta_{2a}(-z-\frac{2\nu}{k})
  \Bigr\} .
\]
In this case, the each summand becomes an eigenstate of
$ \widehat{T} $ with eigenvalue
$ e^{2 \pi \mathrm{i} \cdot 2c/k} $.
By substituting
$ z= (-2\xi-\nu+c+a)/2k $, one obtains matrix elements.
Remark that the difficulty for even-$k$   is due to
Lemma~\ref{lem:sig_sig}.



\section*{Acknowledgement}

The author would like to thank Miki Wadati for his
kind interests in this work.
Thanks are also  to E.~K.~Sklyanin
and  J.~Suzuki
for stimulating and useful discussions.
This work is supported in part by Grand-in-Aid for Encouragement of
Young Scientists
and for Scientific Research on Priority Areas
from
the Ministry of Education, Science and Culture.


\newpage

\begin{thebibliography}{10}

\bibitem{ShiUen92}
Y.~Shibukawa and K.~Ueno,
\newblock Lett. Math. Phys. \textbf{25}, 239 (1992).

\bibitem{FelPas94}
G.~Felder and V.~Pasquier,
\newblock Lett. Math. Phys. \textbf{32}, 167 (1994).

\bibitem{Chere84}
I.~Cherednik,
\newblock Theor. Math. Phys. \textbf{61}, 35 (1984).

\bibitem{Skl88}
E.~K. Sklyanin,
\newblock J. Phys. A: Math. Gen. \textbf{21}, 2375 (1988).

\bibitem{WhittWatso27}
E.~T. Whittaker and G.~N. Watson,
\newblock {\em A Course of Modern Analysis},
\newblock Cambridge Univ. Press, Cambridge,
  1927.

\bibitem{Bela81}
A.~A. Belavin,
\newblock Nucl. Phys.  \textbf{B180}, 189 (1981).

\bibitem{Gaud88}
M.~Gaudin,
\newblock J. Phys. France \textbf{49}, 1857 (1988).

\bibitem{Hikam95d}
K.~Hikami,
\newblock J. Phys. Soc. Jpn. \textbf{64} (1995), in press.

\bibitem{Bax82}
R.~J. Baxter,
\newblock {\em Exactly Solved Models in Statistical Mechanics},
\newblock Academic Press, London, 1982.

\bibitem{Chere82}
I.~Cherednik,
\newblock Sov. J. Nucl. Phys. \textbf{36}, 320 (1982).

\bibitem{Bovie83}
A.~Bovier,
\newblock J. Math. Phys. \textbf{24}, 631 (1983).

\bibitem{Trac85}
C.~A. Tracy,
\newblock Physica \textbf{16D}, 203 (1985).

\bibitem{InamKonn94}
T.~Inami and H.~Konno,
\newblock J. Phys. A: Math. Gen. \textbf{27}, L913 (1994).

\bibitem{VegaGonz94b}
H.~J. de~Vega and A.~Gonz{\'a}lez-Ruiz,
\newblock J. Phys. A: Math. Gen. \textbf{27}, 6129 (1994).

\bibitem{FanHouShiYan95}
H.~Fan, B.-Y. Hou, K.-J. Shi, and Z.-X. Yang,
\newblock Phys. Lett.  \textbf{A200}, 109 (1995).

\bibitem{KuliSkly91}
P.~P. Kulish and E.~K. Sklyanin,
\newblock J. Phys. A: Math. Gen. \textbf{24}, L435 (1991).


\end{thebibliography}
\end{document}